# A photon counting reconstructive spectrometer combining metasurfaces and superconducting nanowire single-photon detectors

Jingyuan Zheng[1,†], You Xiao[1,†], Mingzhong Hu[1], Yuchen Zhao[1], Hao Li[2], Lixing You[2], Xue Feng[1,3], Fang Liu[1,3], Kaiyu Cui[1,3], Yidong Huang[1,3,4] and Wei Zhang[1,3,4,*]

[1]*Beijing National Research Center for Information Science and Technology (BNRist), Beijing Innovation Center for Future Chips, Department of Electronic Engineering, Tsinghua University, Beijing 100084, China*
[2]*State Key Laboratory of Functional Materials for Informatics, Shanghai Institute of Microsystem and Information Technology, Chinese Academy of Sciences, Shanghai 200050, China*
[3] *Frontier Science Center for Quantum Information, Beijing 100084, China*
[4]*Beijing Academy of Quantum Information Sciences, Beijing 100193, China*
* zwei@tsinghua.edu.cn
† These authors contributed equally: Jingyuan Zheng, You Xiao

**Abstract**

Faint light spectroscopy has many important applications such as fluorescence spectroscopy, lidar and astronomical observations. However, long measurement time limit its application on real-time measurement. In this work, a photon counting reconstructive spectrometer combining metasurfaces and superconducting nanowire single photon detectors (SNSPDs) was proposed. A prototype device was fabricated on a silicon on isolator (SOI) substrate, and its performance was characterized. Experiment results show that this device support spectral reconstruction of mono-color lights with a resolution of 2 nm in the wavelength region of 1500 nm ~ 1600 nm. The detection efficiency of this device is 1.4% ~ 3.2% in this wavelength region. The measurement time required by this photon counting reconstructive spectrometer was also investigated experimentally, showing its potential to be applied in the scenarios requiring real-time measurement.

**Introduction**

Since optical spectrum has rich information of the interaction between light and matter, spectrometers are widely applied in various fields of science and technology [1]. In many applications, such as fluorescence spectroscopy [2], spectral lidar [3, 4] and astronomical observation [5], the measured light is at single-photon-level. Usually, the spectrometer for such a faint light is realized by a monochromator and a single photon detector [3, 6, 7]. The spectral information is obtained by photon counting when the narrow filter of the monochromator is set at different wavelengths. Long measurement time is required, partly due to its low photon utilization, since only the photons in the narrow filter bandwidth are detected under a specific setting of the monochromator, while, most photons out of the filter are simply rejected. Besides, the scanning process of the monochromator also contributes to the long measurement time. In some applications, high performance image sensors are used to avoid the scanning process [8, 9]. However, since each pixel on the sensor also only detects photons in a very narrow band, long measurement time is also required to achieve a signal to noise ratio high enough. Hence, spectral sensing of ultra-faint light is difficult to be applied in the scenarios requiring real-time measurement. Besides, long measurement time also means more noise counts would be recorded during the measurement.

In recent years, reconstructive spectrometers based on compressive sensing developed rapidly [10]. They are based on a detector array and optical filters with different complex transmission spectra. The spectral response of the detectors on the array are modulated by these filters. The spectral information of the incident light can be reconstructed by the measurement results of the detectors and their spectral responses. Usually, the number of the detectors is far smaller than the sample point number of the spectrum, hence, algorithms of compressive sensing are used to reconstructed the spectrum. The commercial image sensors provide simple and low-cost solutions for the detector arrays. Hence, many works focused on the materials and structures to realize the optical filters, including quantum dots [11], nanowires [12], and metasurfaces [13, 14]. Recently, a real-time ultra-spectral image sensor with over one hundred thousand pixels based on an image sensor and metasurfaces was reported, showing it is potential on developing minimized devices for spectral imaging [15]. We'd like to point out that the principle of reconstructive spectrometers also can be applied on the scenarios of ultra-faint light. The optical filters in the reconstructive spectrometer have broadband complex transmission spectra. It would lead to much higher photon utilization comparing with the schemes based on monochromators or other narrow band optical filters. Besides, it supports "snapshot" measurement, without scanning process since all the detectors take photon counting simultaneously. Hence, the measurement time would be highly reduced by a reconstructive spectrometer of ultra-faint light based on photon counting. The key point of such a device is the way to combine the single photon detectors and the optical filters.

Among all the techniques of single photon detection, the superconducting nanowire single photon detector (SNSPD) is a good candidate to develop on-chip spectrometer of ultra-faint light [16]. Some intrinsic characteristics of SNSPDs have been applied to realize spectral sensing and analysis, such as the spectral response variation of a SNSPD under different bias currents [17] and the relation between the amplitudes of single photon events and the wavelengths of detected photons [18]. However, such intrinsic characteristics could not provide enough freedoms to modulate the spectral responses of SNSPDs and support high performance spectral sensing. On the other hand, as an ultra-thin metal wire, the SNSPD is convenient to be integrated on a variety of substrates and photonic structures, showing good compatibility with various fabrication processes [19-21]. Two traditional spectrometer devices have been reported, in which an arrayed waveguide grating [22] and a Rowland circle [23] were used to realize the spectral response modulation of the SNSPD, respectively. Recently, an in-plane disordered scattering structure was introduced to modulate the spectral response of SNSPDs, realizing a single-photon-sensitivity reconstructive spectrometer [24]. However, the detection efficiencies achieved in these works were quite low. Especially, the requirement of long measurement time is still an unsolved problem, even for the reconstructive spectrometer.

In this work, we proposed a photon counting reconstructive spectrometer combining SNSPDs and metasurfaces. In the device, the SNSPDs were fabricated on the metasurfaces with different structure parameters, which modulated the spectral responses of the SNSPDs, supporting the spectral reconstruction by compressive sensing. A prototype device was fabricated and characterized, showing its performance on spectral measurement and reconstruction of ultra-faint light, and its potential to be applied in the scenarios requiring real-time measurement.

**Methods**

The sketch of the proposed device is shown in Fig. 1(a). It realizes on a substrate with high index thin film layer. A metasurface array is fabricated on the thin film layer by etching periodically arranged micro-holes. The metasurface array is divided into many regions, each of them has different structure parameters, such as the period and the size of the micro-holes. In each region, a SNSPD is fabricated in the spaces among micro-holes. Hence, the spectral responses of these SNSPDs are modulated by the corresponding metasurfaces, showing complex profiles with low correlation. By this way, each region of the metasurface array become a spectral sensing unit with a specific measurement basis.

Considering that an ultra-faint light with a photon flux density of $f(\lambda)$ illuminates the devices with $N$ spectral sensing units, $\lambda$ is the wavelength. $M$ evenly spaced sample points are considered in the spectral region for the measurement, $\lambda_j$ is the $j$-th sample point and $\delta\lambda$ is the interval between two sample points. The photon count number of the $i$-th sensing unit in a measurement is denoted by $c_i$, which can be expressed as

$$\begin{bmatrix} c_1 \\ c_2 \\ \vdots \\ c_N \end{bmatrix} = \begin{bmatrix} r_1(\lambda_1) & r_1(\lambda_2) & \cdots & r_1(\lambda_M) \\ r_2(\lambda_1) & r_2(\lambda_2) & \cdots & r_2(\lambda_M) \\ \vdots & \vdots & \ddots & \vdots \\ r_N(\lambda_1) & r_N(\lambda_2) & \cdots & r_N(\lambda_M) \end{bmatrix} \begin{bmatrix} f(\lambda_1)\delta\lambda T \\ f(\lambda_2)\delta\lambda T \\ \vdots \\ f(\lambda_M)\delta\lambda T \end{bmatrix} \quad (1)$$

Where $T$ is the measurement time, i. e. the time of photon counting. $r_i(\lambda_j)$ is the spectral response of the SNSPD in the $i$-th spectral sensing unit and at the $j$-th spectral sample point. They would be measured at the calibration process of the device. Usually, the sample point number $M$ is far larger than the spectral sensing unit number $N$, hence, Eq. (1) is an underdetermined linear equation group of $f(\lambda_j)$. The optimal solution of Eq. (1) can be calculated under certain constraints or prior information, which actually reconstructs the spectrum of input faint light through compressed sensing algorithms [25].

Based on the proposed scheme, we designed and fabricated a prototype device. It was fabricated on a silicon on isolator (SOI) substrate with a silicon layer of 340-nm in thickness. Niobium nitride (NbN) was deposited on the silicon layer. Then, the NbN nanowires were patterned via electron-beam lithography (EBL) and etched by reactive-ion etching (RIE), realizing the fabrication of SNSPDs. The electrodes of the nanowires were also made by the NbN film using UV lithography and RIE process. Finally, the metasurface array was defined by EBL and etched by the inductively coupled plasma reactive-ion etching (ICP-RIE) at the silicon layer. Fig. 1(b) and (c) are scanning-electron micrograph (SEM) pictures to show the full view of the device and details of some sensing units on it. The device was mounted in a cryostat under a temperature of 2.1K and coupled with an optical fiber for optical input.

We measured the spectral responses of all the spectral sensing units on the device, obtaining $r_i(\lambda_j)$ in Eq. (1). The wavelength region of the measurement was 1500 nm ~ 1600 nm, with 101 sample points ($M = 101$, $\delta\lambda = 1$ nm). The light for the measurement is provided by a tunable laser (Santec TSL-510). Details of the experimental setup and processes are introduced in the Supplementary Materials. The measured spectral responses of all the sensing units are shown in Fig. 1(d) and it can be recognized that the device has 9 spectral sensing units in operation. All of them have complex spectral responses, which are different from each other, showing low correlation between any two spectral measurement bases of the device. We also fabricated a sample without metasurfaces, on which only the SNSPDs were fabricated. The spectral responses of SNSPDs in this sample were also measured for comparison and the results are shown in Supplementary Materials, which indicates that the complex spectral responses of the device shown in Fig.1 (d) are due to the modulations of the metasurfaces with different structure parameters. The total efficiency of this device is calculated by summing the spectral responses of all the sensing units, which is shown in Fig. 1(e) and it is in a range of 1.4% ~ 3.2% in the wavelength region of the measurement, which is comparable to the performance of SNSPDs on similar substrates without reflective or resonance structures [26].

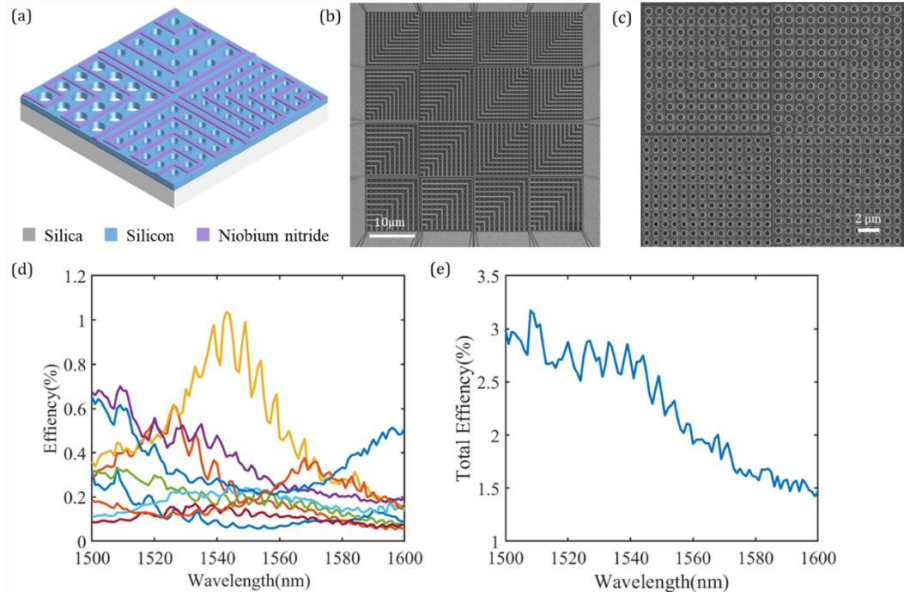

Fig.1 Sketch of the proposed photon counting reconstructive spectrometer and SEM pictures of the prototype device. (a) Sketch of the spectrometer, which is a metasurface array with SNSPDs in different regions; (b) The SEM picture of the full view of the device; (c) The SEM picture of 4 spectral sensing units in the device. (d) Spectral responses of all the spectral sensing units. (e) The spectrum of total detection efficiency of this device.

**Results**

*Spectrometer characterizations*

The performance of the prototype device at telecom band (1500 nm~1600 nm) was demonstrated experimentally. Fig. 2(a) shows the spectral reconstruction results when attenuated monochromatic lights at different wavelengths were measured by the device, which were provided by the same tunable laser for the device calibration. It shows that the wavelength of the monochromatic light is reconstructed successfully in the whole wavelength region of the measurement. To show the wavelength resolution of the device, two attenuated monochromatic lights with a wavelength difference of 2 nm were generated at different wavelengths by two tunable lasers and measured by the device. The spectral reconstruction results are shown in Fig. 2(b). It suggests that the two monochromatic lights can be discriminated at all the wavelength setting, indicating that the device supports a wavelength resolution of 2 nm over the wavelength region of 100 nm.

Since the SNSPDs have broad wavelength band of single photon detection, and the metasurfaces also can provide broadband spectral response modulations on the SNSPDs, it can be expected that the device can support the spectral measurement and reconstruction in a much broader operation band. Hence, we also calibrate the device in a wavelength region of 1350 nm ~ 1629 nm and explore its performance as a spectrometer in such a broad band. The results are shown in the Supplementary Materials.

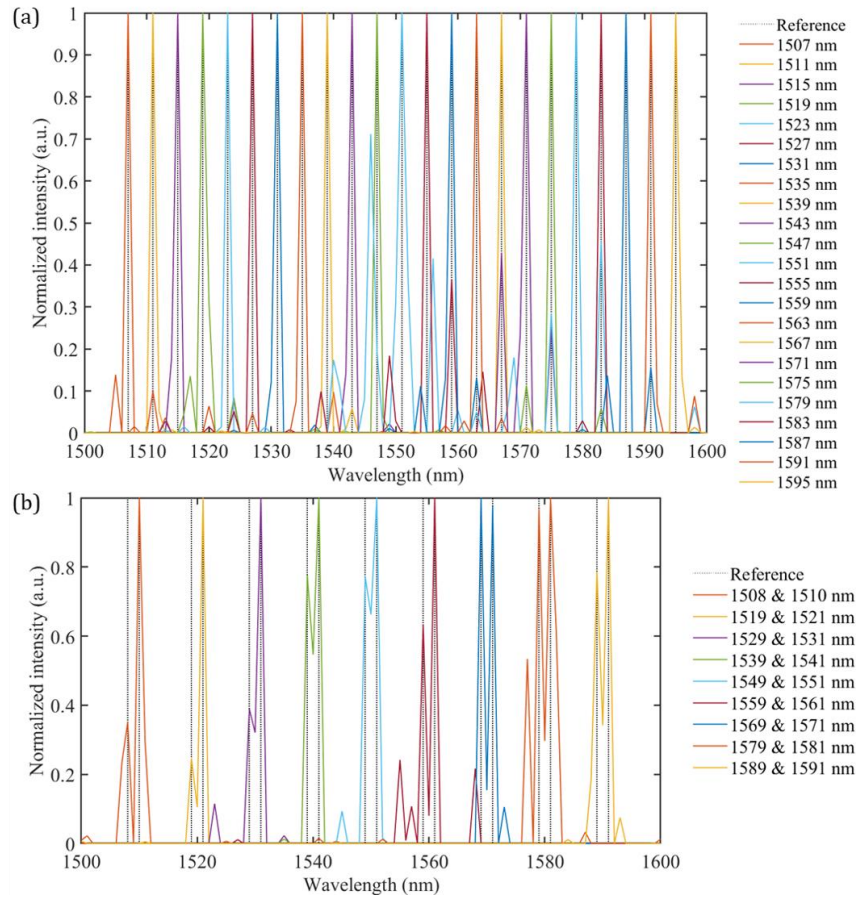

Fig.2. Spectrometer characterizations from 1500 nm to 1600 nm. (a) Reconstruction results of attenuated monochromatic lights at different wavelengths. (b) Reconstruction results of two monochromatic lights with a wavelength interval of 2 nm at different wavelength settings.

*Measurement time required by the photon counting reconstructive spectrometer*

The performance of spectral reconstruction of this device depends on the photon counts it records. The single photon events recorded by each spectral sensing units satisfy Poisson statistics, hence, the photon count number of the $i$-th sensing unit ($c_i$) should satisfy Gaussian distribution according to the central limit theorem. The standard deviation of the Gaussian distribution is $\sqrt{k}$, if the average value of $c_i$ is $k$ under a measurement time of $T$. It can be expected that the relative uncertainty of the measured ci would decreases under an increasing $T$, which is helpful to improve the quality of spectral reconstruction. On the other hand, in principle the scheme of reconstructive spectrometer is helpful to save measurement time by enhancing the photon utilization in the measurement of photon counting. It is because that the sensing units in the device have complex spectral responses, which could detect more photons comparing with the traditional schemes based on narrow optical filters. Hence, a natural question is what is the proper measurement time required by the photon counting reconstructive spectrometer?

It was investigated by comparing the spectral reconstruction quality of this device under different measurement times. In the experiment, the faint light to be measured was an attenuated broadband light with a Gaussian profile in its spectrum. Its center-wavelength and full width at half maximum (FWHM) were 1540 nm and 12 nm, respectively. The single photon events of all the sensing units on the device were recorded in 60 s, showing a total photon count rate of about 29 MHz. For a specific measurement time $T$ ($T < 0.3$ s), one hundred of independent time bins with a width of $T$ were selected in the 60 s. The recorded single photon events in each time bin were counted to take the spectral reconstruction. The quality of the reconstructed spectrum was evaluated by the root mean square error (RMSE), which is defined as $RMSE = \sqrt{\frac{1}{M}\sum_{j=1}^{M}[f(\lambda_j) - f_0(\lambda_j)]^2}$, where $f(\lambda_j)$ and $f_0(\lambda_j)$ are the reconstructed spectrum and the actual spectrum of the broadband light, respectively. The profile of $f_0(\lambda_j)$ was measured by an optical spectrum analyzer (OSA) before it was attenuated. The statistics of the RMSEs obtained by the data of all these time bins were calculated in Fig. 3. Each point in Fig. 3 shows the average and standard deviation of the RMSEs for a specific measurement time $T$, which is the bottom x-axis of Fig. 3. The corresponding average photon count number also indicates at the top x-axis of Fig. 3. The comparisons between the reconstructed spectrum and the actual spectrum under different RMSE level are also shown in Fig. 3. Both the average and the standard deviation of the RMSE decrease under increasing measurement time, i.e. the number of single photon events the device recorded. On the other hand, the main components of the actual spectrum could be reconstructed quite well when the RMSE is smaller than 0.1. Hence, a successful spectral reconstruction can be defined by RMSE < 0.1 in this experiment. A reasonable measurement time should ensure a high possibility that the spectrum reconstruction of a measurement is successful. If this condition is defined by that the average RMSE should be lower than 0.1 by two standard deviations, the shortest measurement time is about 30 ms. To show the ability of this device on fast measurement, a wavelength sweeping attenuated monochromatic light were measured and results are introduced in the Supplementary Materials. These results show that the device has the potential to be applied in the scenarios requiring fast spectral measurement and analysis on single-photon-level faint light.

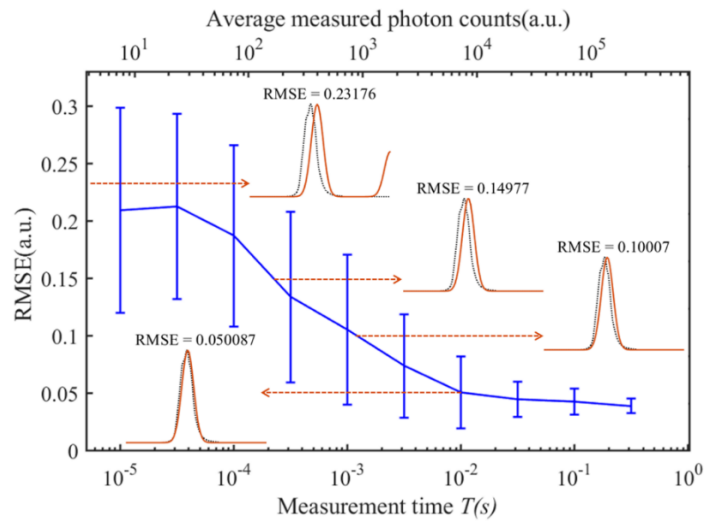

Fig.3 The statistics of the RMSE of the reconstructed spectrum for a specific broadband light under different measurement time $T$. The up x-axis is the average photon counts recorded by the device, which are calculated by $T$ and the total photon count rate. Error bar is the standard deviation of the RMSE. Inset images are the typical reconstruction results under the RMSE of 0.23, 0.15, 0.10 and 0.05.

## Conclusions

In this work, we proposed a photon counting reconstructive spectrometer combining metasurfaces and SNSPDs. The SNSPDs were fabricated among the micro-holes of metasurfaces with different structure parameters. Their spectral responses were modulated by corresponding metasurfaces, realizing complex spectral measurement bases with low correlation. The spectrum of the incident faint light is reconstructed by compressive sensing. We fabricated a prototype device on a SOI substrate, and characterized its performance in the wavelength region of 1500nm~1600nm. Experiment results show that this device support spectral reconstruction of mono-color lights with a resolution of 2 nm in the wavelength region of 1500 nm ~ 1600 nm. The detection efficiency of this device is 1.4% ~ 3.2% in this wavelength region. It is worth noting that the device could operate in a much broader wavelength region, which are shown in the Supplementary Materials.

The measurement time required by this photon counting reconstructive spectrometer was also investigated experimentally, showing its potential to be applied in the scenarios requiring real-time measurement. It can be expected that better performance could be expected if the detection efficiency could be enhanced by improving the fabrication process [27], the substrate [28], and the SNSPD structures [29]. Besides, the optimization of spectral measurement bases and spectral reconstruction algorithm would also highly improve the performance of the proposed device.

**Supplementary Information**

Additional file 1. See the supplementary material for spectral response calibration details, spectral reconstruction method, spectral responses of SNSPDs in the sample without metasurfaces, spectrometer characterizations at the wavelength region from 1350 nm to 1629 nm and fast spectral measurements.

**Authors' contributions**

Jingyuan Zheng and You Xiao contributed equally to this work. Wei Zhang and Jingyuan Zheng proposed the scheme. Jingyuan Zheng and You Xiao designed the device. You Xiao fabricated the SNSPDs on the device. Jingyuan Zheng, You Xiao and Mingzhong Hu performed experiments and analyzed data. Jingyuan Zheng and Yuchen Zhao took the theoretical analysis. Jingyuan Zheng and Wei Zhang wrote the manuscript. Yidong Huang revised the manuscript and supervised the project. Hao Li, Lixing You, Xue Feng, Fang Liu and Kaiyu Cui contributed to discussion of this study and the revision of the manuscript. All authors read and approved the final manuscript.

**Acknowledgements**

We thank PHOTEC for help with the cryostat assembly and testing.

**Funding**

National Key R&D Program of China (2017YFA0303704), Beijing Natural Science Foundation (BNSF) (Z180012), National Natural Science Foundation of China (NSFC) (61875101, 91750206), and the Tsinghua University Initiative Scientific Research Program.

**Availability of data and materials**

The datasets used and analyzed during the current study are available from the corresponding author on reasonable request.

# Supplementary Materials for

# A photon counting reconstructive spectrometer combining metasurfaces and superconducting nanowire single-photon detectors: supplemental document

**Jingyuan Zheng[1],[†], You Xiao[1],[†], Mingzhong Hu[1], Yuchen Zhao[1], Hao Li[2], Lixing You[2], Xue Feng[1,3], Fang Liu[1,3], Kaiyu Cui[1,3], Yidong Huang[1,3,4] and Wei Zhang[1,3,4], ***

[1]*Beijing National Research Center for Information Science and Technology (BNRist), Beijing Innovation Center for Future Chips, Department of Electronic Engineering, Tsinghua University, Beijing 100084, China*

[2]*State Key Laboratory of Functional Materials for Informatics, Shanghai Institute of Microsystem and Information Technology, Chinese Academy of Sciences, Shanghai 200050, China*

[3] *Frontier Science Center for Quantum Information, Beijing 100084, China*

[4]*Beijing Academy of Quantum Information Sciences, Beijing 100193, China*

* *zwei@tsinghua.edu.cn*

*† These authors contributed equally: Jingyuan Zheng, You Xiao*

**The spectral response calibration of the prototype device**

The prototype device was mounted in a cryostat under a temperature of 2.1K. The ultra-faint light to be measured was coupled to the device through a single mode fiber. Supplementary Fig. 1 shows the experiment setup of the spectral response calibration of the device. The attenuated monochromatic light at each sample point was generated from a tunable laser (Santec TSL-510F) and attenuated by a pair of tunable attenuators (KEYSIGHT 81571A). Then, it connected to the input port of an optical switch ((Thorlabs OSW12-1310E)). The optical switch had two output ports. One of them connected to a calibrated optical power meter (Thorlabs S154C and PM100D), the other connected to the single mode fiber coupling the light to the device. The SNSPD of each spectral sensing unit in the device was supported by an independent circuit, which provided its bias current. The circuit also coupled out its output signal to a channel of a streaming time-to-digital converters (Swabian Instruments TimeTagger), which recorded the photon counts of this SNSPD[1].

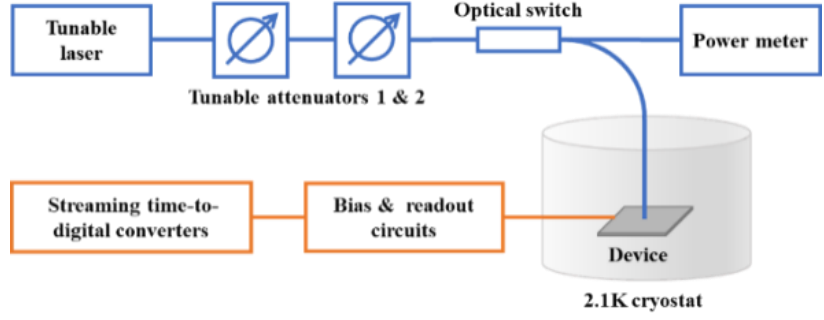

Supplementary Fig. 1: Schematic diagram of the experimental setup of the spectral response calibration.

The photon flux of the attenuated monochromatic light could be calculated by the measured power and the settings of the two attenuators. To reduce the uncertainty of the calibration process, we measured the detection efficiency of the SNSPD in each sensing units at a specific frequency sample point $\lambda_j$ following these steps:

1) The light was directed to the calibrated power meter by the optical switch. The power of the tunable laser was set to $P$, attenuations of the two attenuators were both set to 0dB. The result of the power meter was recorded as $P_0(\lambda_j)$.

2) The attenuations of the two attenuators were set to $a_1$ dB and $0$ dB, respectively. The result of the power meter was recorded as $P_1(\lambda_j)$.

3) The attenuations of the two attenuators were set to $0$ dB and $a_2$ dB, respectively. The result of the power meter was recorded as $P_2(\lambda_j)$.

4) The light was directed to the single mode fiber coupling the light to the device by the optical switch. The attenuations of the two attenuators were set to $a_1$ dB and $a_2$ dB, respectively. Then the photon count rate of the SNSPD in the $i$-th spectral sensing unit are recorded as $PCR_i(\lambda_j)$.

5) The detection efficiency of the SNSPD in the $i$-th spectral sensing unit at the $j$-th wavelength sample point $r_i(\lambda_j)$ was calculated by

$$r_i(\lambda_j) = PCR_i(\lambda_j) \frac{P_0(\lambda_j)}{P_1(\lambda_j)P_2(\lambda_j)} \frac{hc}{\lambda_j} \qquad (S1)$$

Here, $h$ is Planck constant and $c$ is the speed of light.

The spectral responses of all the sensing units in the device were obtained when their detection efficiencies at all the wavelength sample points were measured. The total efficiency spectrum of the device was calculated by summing the spectral responses of all the sensing units.

**Compressed sensing for spectral reconstructions**

As the principle mentioned in Methods of the text, the spectral measurement can be described as $c = Rf$, in which $c$ is the vector of photon count number obtained by all the spectral sensing units, $R$ is the spectral response matrix calibrated through the measurement introduced in S1 and $f$ is the spectrum to be reconstructed. Since the number of sensing units is usually less than the number of wavelength sample points, it is an under-determined equation. When the prior information of $f$ is known, the equation is transformed into an optimization problem[2]:

$$\min_{f \geq 0}\{\|c - Rf\|_2 + \tau P(f)\} \tag{S2}$$

Here, $\|\cdot\|_2$ represents the second norm, $P(f)$ is the mathematical expression of the prior information and $\tau$ is the regularization parameter, which controls the influence of prior information during the optimization process. Many algorithms can be used to solve such optimization problems, such as convex optimization.

**Spectrometer characterizations at the wavelength region of 1350 nm ~ 1629 nm**

The attenuated monochromatic light at the wavelength region from 1350 nm to 1629 nm was generated by a tunable laser (Santec TSL-210F) and attenuated by cascaded tunable attenuators. The total number of wavelength sample points was 93, with a wavelength interval of 3 nm. According to the calibration method mentioned above, the spectral responses of these sensing units in this wavelength region were measured, which are shown in Supplementary Fig. 2(a). The total efficiency spectrum of the device is shown in Supplementary Fig. 2(b) and the total efficiency is in a range of 1.1% and 4.4%.

The performance of spectral sensing and reconstruction at this wavelength region was also demonstrated experimentally. Supplementary Fig. 2(c) shows the spectral reconstruction results when attenuated monochromatic lights at different wavelengths were measured by the device. The central wavelength of the monochromatic light can be reconstructed successfully over the whole measurement wavelength region. Supplementary Fig. 2(d) is a typical spectral reconstruction result when an attenuated broadband light was measured, which was provided by amplified spontaneous emission (ASE) source based on a piece of erbium-doped fiber (EDF). It clearly shows that the main spectral components of the broadband light can be obtained by this device. To show the wavelength resolution of this device, two attenuated monochromatic lights with a wavelength difference of 6 nm, 9 nm, 12 nm and 15 nm were provided near 1550nm. The spectral reconstruction results of this device are shown in Supplementary Fig. 2(e)-(j), respectively and the two monochromatic lights can be discriminated in all the cases.

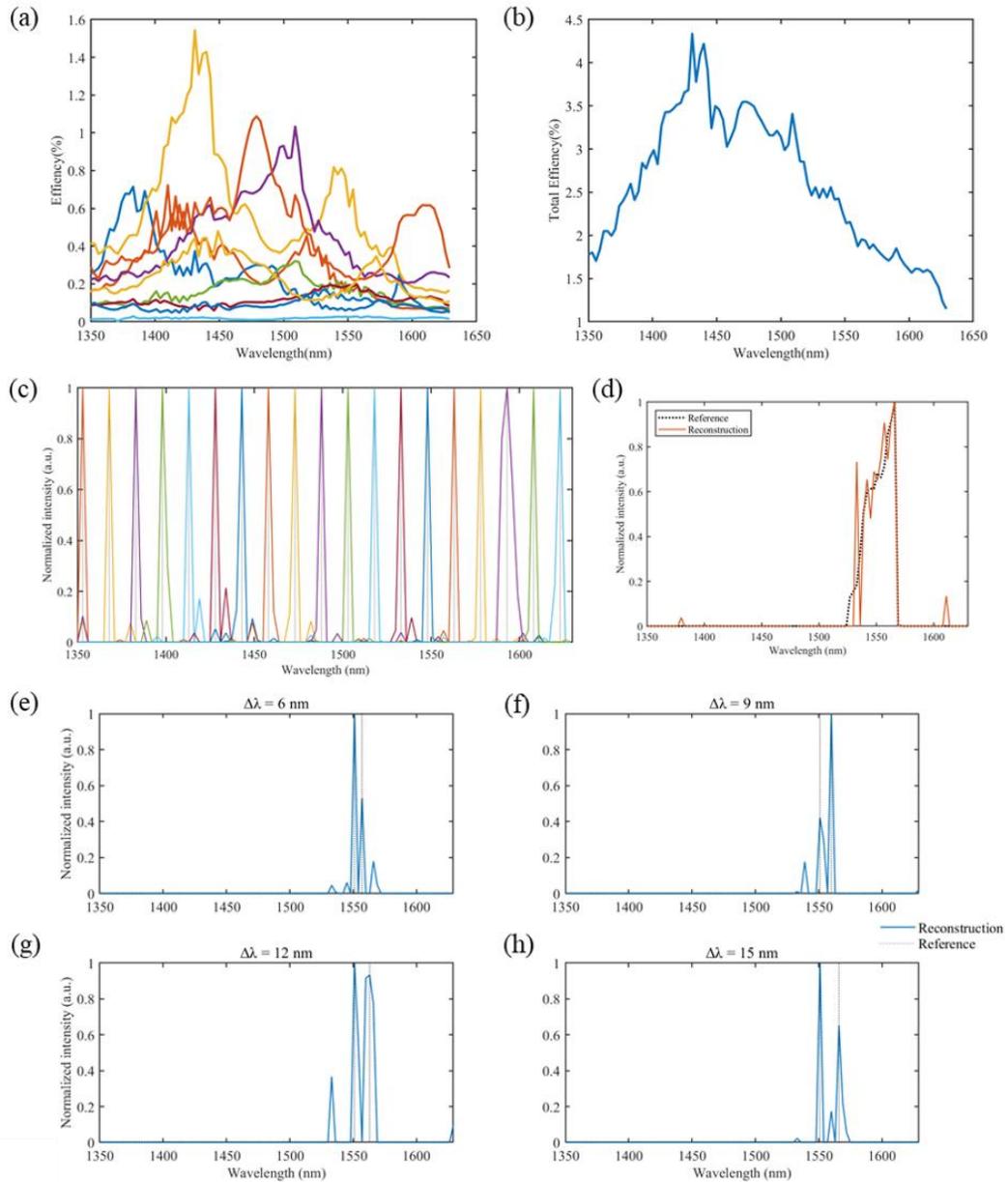

Supplementary Fig. 2: Spectrometer characterizations from 1350nm to 1629nm. (a) Spectral responses of each spectral sensing unit. (b) Total detection efficiency spectrum of the device. (c) Reconstruction results of attenuated monochromatic lights at different wavelengths; (d) A typical reconstruction result of an attenuated broadband light provided by an EDF ASE source; (e)-(j) Reconstruction results of two monochromatic lights near 1550nm with a wavelength interval of 6nm, 9nm, 12nm and 15nm, respectively.

**Spectral responses of SNSPDs in the sample without metasurfaces**

To show the effect of metasurfaces in the proposed photon counting reconstructive spectrometer, we also fabricated a sample without metasurfaces. Only the SNSPDs were fabricated on the sample with the same structure parameters with those of the prototype device. We measured the spectral responses of SNSPDs in this sample for comparison. Supplementary Fig. 3(a) is the SEM picture of the sample. The measured spectral responses of different SNSPDs are shown in Supplementary Fig. 3(b). Comparing with the results of the prototype device shown in Supplementary Fig. 2(a), it is obvious that the spectral responses of this sample are quite similar with each other, which indicates that the complex spectral responses of the prototype device is brought by the metasurfaces, which is fundamental for the spectral reconstruction performance. Supplementary Fig. 3(c) shows the total efficiency of the sample without metasurfaces. It can be found that the total efficiency of the sample is smaller than that of the prototype device. It indicates that the enhancement of local light field intensity introduced by the metasurfaces is also beneficial to enhance the absorption of SNSPDs, leading to an improvement of the detection efficiencies. Considering that only 9 SNSPDs were active in the prototype device, while 16 SNSPDs of the sample without metasurfaces were all active, the effect of efficiency improvement of the metasurfaces is quite obvious.

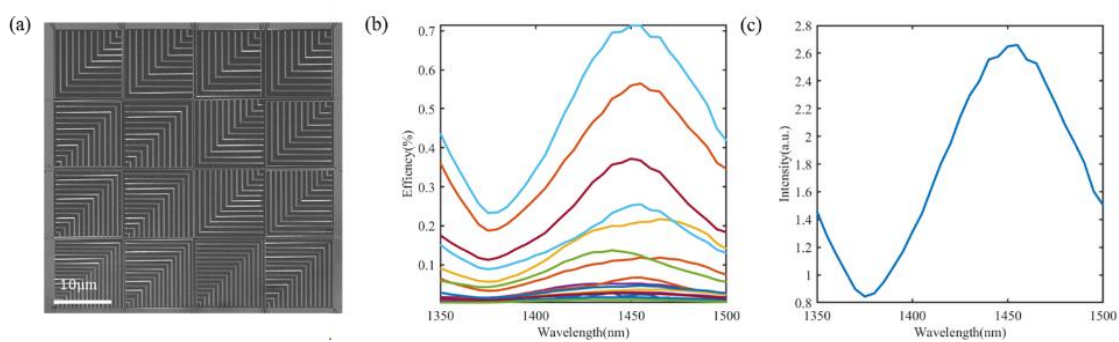

Supplementary Fig. 3: Spectral responses of SNSPDs in the sample without metasurfaces. (a) The SEM picture of the sample. (b) Spectral responses of the SNSPDs. (c) Total efficiency of the sample

**Fast spectral measurement of the sweeping monochromatic faint light**

To show the ability of this device on fast measurement, an attenuated monochromatic light with its wavelength sweeping from 1547 nm to 1554 nm was generated by the tunable laser (Santec TSL510F). Its wavelength varied with increasing time as shown in Supplementary Fig. 4(a). It shows that the wavelength would increase by 1 nm after a time-interval of $T_w$, which could be adjusted by the setting of the tunable laser. The photon flux of the attenuated monochromatic light was set to support a counting rate of about 17.55 MHz detected by the device. Each time-interval of $T_w$ was divided into 10 time-bins, and the spectral reconstructions were achieved based on the data of single photon events recorded in these time-bins. Hence, the dynamic variation of the spectrum was measured under a sampling time interval of $\delta T_w$, which was 1/10 of $T_w$.

In the experiment, $T_w$ was set to 1 s, 0.5 s and 0.2 s, respectively, the corresponding $\delta T_w$ is 100 ms, 50 ms and 20 ms, respectively. The results of the dynamic spectral reconstruction spectra under these settings of $T_w$ and $\delta T_w$ are shown in Supplementary Fig. 4(b)-(d), respectively. In these figures, the x- and y-axes represent time and wavelength, respectively. The color bars show the normalized intensity of the reconstructed spectra. These results show that the reconstructed spectra agree with the laser setting quite well, showing the temporal wavelength variation of the attenuated monochromatic light in all the cases. On the other hand, the quality of spectral reconstruction shows a little decline when $\delta T_w$ reduces from 100 ms to 20 ms. It is because that smaller $\delta T_w$ leads to smaller photon counts of the SNSPDs in the spectral sensing units. Higher uncertainty could be expected under a smaller $\delta T_w$ due to the Gaussian statistics of the photon counting, which would impact the quality of the spectral reconstruction. This experiment shows that the device can support a fast spectral reconstruction with a measurement time of 10 ms ~ 100 ms, under a photon counting rate of ~ 10 MHz.

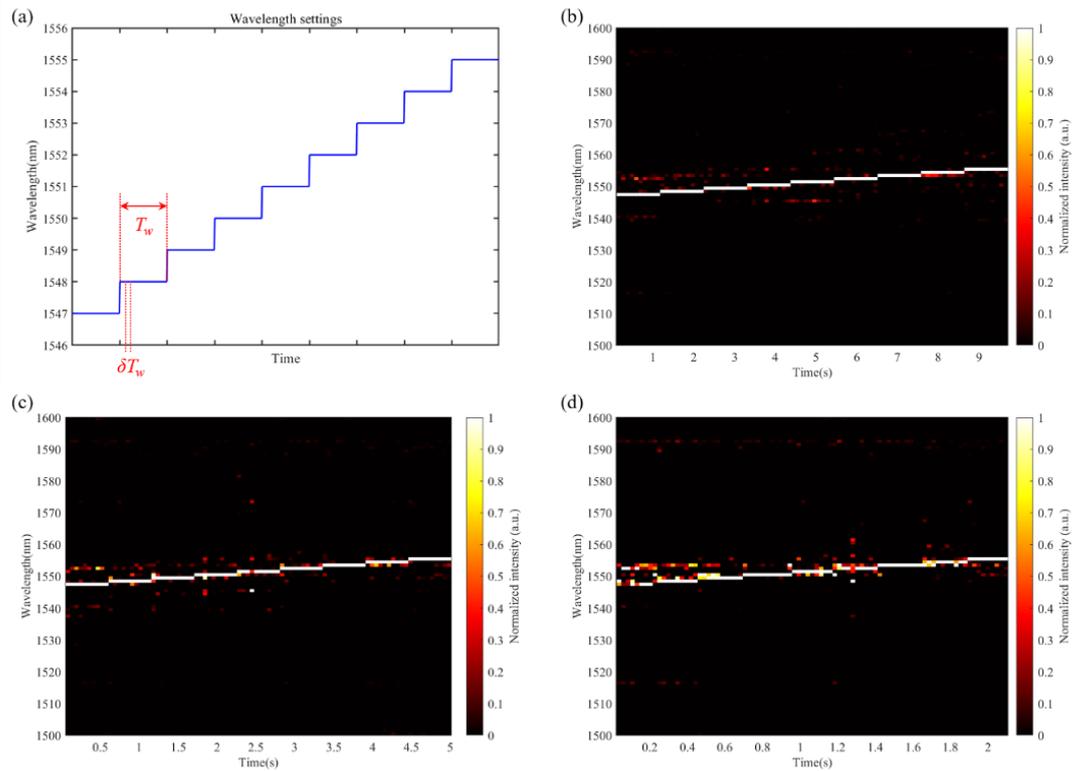

Supplementary Fig. 4: Experiment results of fast spectral measurement. (a) Schematic sketch of the temporal wavelength variation of the attenuated monochromatic light. (b)-(d) The reconstructed spectra measured under different settings of $T_w$ and $\delta T_w$ when the wavelength of the attenuated monochromatic light sweeping from 1547 nm to 1554 nm. (b) $T_w$ =1s and $\delta T_w$ = 100ms; (c) $T_w$ =0.5 s and $\delta T_w$ = 50 ms; (d) $T_w$ =0.2 s and $\delta T_w$ = 20 ms.

**Supplementary References:**